\documentclass[prb,aps,twocolumn,showpacs]{revtex4}
\usepackage{epsfig}
\usepackage{amsmath}
\begin{document}

\title{Effect of linear density of states on the quasi-particle
dynamics and small electron-phonon coupling in graphite}

\author{C. S. Leem$^1$, B. J. Kim$^2$, Chul Kim$^1$, S. R. Park$^1$,
T. Ohta$^3$, A. Bostwick$^3$, E. Rotenberg$^3$, H. -D. Kim$^4$, M.
K. Kim$^1$, H. J. Choi$^1$, and C. Kim$^{1,*}$}

\affiliation{$^1$Institute of Physics and Applied Physics, Yonsei
University, Seoul, Korea}

\affiliation{$^2$School of Physics and Center for Strongly
Correlated Materials Research, Seoul National University, Seoul,
Korea}

\affiliation{$^3$Advanced Light Source, Lawrence Berkeley National
Laboratory, Berkeley, California 94720, USA}

\affiliation{$^4$Pohang Accelerator Laboratory, Pohang 790-784,
Korea}

\date{\today}

\begin{abstract}
We obtained the spectral function of very high quality natural
graphite single crystals using angle resolved photoelectron
spectroscopy (ARPES). A clear separation of non-bonding and bonding
bands and asymmetric lineshape are observed. The asymmetric
lineshapes are well accounted for by the finite photoelectron escape
depth and the band structure. The extracted width of the spectral
function (inverse of the photohole life time) near the $K$ point is,
beyond the maximum phonon energy, approximately proportional to the
energy as expected from the linear density of states near the Fermi
energy. The upper bound for the electron-phonon coupling constant is
about 0.2, a much smaller value than the previously reported one.
\end{abstract}
\maketitle

Recent discoveries of novel physical properties in carbon-based
materials such as superconductivity
\cite{Hebard,Tang,Kociak,Weller,Emery} and massless Dirac Fermions
\cite{Novoselov} brought renewed interest in the electronic
structure of graphite\cite{Zhou,Sugawara}. The peculiarity of the
electronic structure of graphite has two aspects: graphite is
extremely two dimensional and is a semi-metal. These facts make it
fundamentally interesting to study how dimensionality affects the
dynamics of the doped carriers and how the carriers in graphite
intercalated compounds (GICs) couple to the mediating bosons. In
fact, there is a long standing issue in regards to the carrier
dynamics in graphite, that is, whether the carriers are
Fermi-liquid-like or not. This question motivates experimental
studies of electronic structures of these materials by using, for
example, angle-resolved photoemission spectroscopy (ARPES) and one
can find a long history in the ARPES studies on graphite single
crystals\cite{Law,Takahashi,Law2,Maeda,Collins,
Strocov,Balasubramanian,Zhou,Sugawara}. In addition, studies of
graphite-related materials such as single\cite{Bostwick} and
bilayer graphene\cite{Ohta} and GICs\cite{Fretigny,Molodtsov} can
be found.

High quality ARPES data from graphite is difficult to obtain despite
graphite's two dimensional, inert nature. The problems associated
with ARPES experiments on graphite are due mostly to difficulty in
proper surface preparation and to some extent to the low quality of
the single crystals. For example, the extreme two-dimensional nature
of graphite inevitably produces small flakes (many are small enough
to be seen only under microscopes) on the cleaved surfaces which
ruins the momentum resolution in ARPES. Such difficulties prevented
one from obtaining good quality data to extract reliable information
on the many-body interactions such as electron-phonon coupling
(EPC). Therefore, unless such difficulties are overcome, reliable
information on many-body interactions can not be extracted from the
data. As a result, the experimental data in regards to the electron
lifetime have been obtained mostly by time-resolved photoelectron
spectroscopy on highly oriented pyrolytic graphite\cite{Xu,Moos}.

Motivated by the renewed interest in the carrier dynamics in
graphite, we have performed ARPES studies on graphite single
crystals. Our goal was to extract reliable quantitative
information on the EPC from ARPES data. To overcome the above
mentioned difficulties, we exploited micro-spot ARPES and high
quality natural graphite single crystals, and were able to obtain
ARPES data with high enough quality to reliably extract
quantitative information on the electron-phonon coupling. In
addition, we successfully applied the finite photoelectron escape
depth and band structure effects in graphite to the analysis of
the ARPES spectral functions for the first time (to our
knowledge). This allowed us to obtain an accurate measure of the
EPC in graphite by using ARPES.

ARPES experiments were performed at the beam line 3A1 of the Pohang
Accelerator Laboratory (preliminary) and 7.0.1 of the Advanced Light
Source (main results).  High quality natural graphite single
crystals with sizes larger than 1 cm were cleaved \emph{ex situ}.
The samples were subsequently introduced to the vacuum chamber and
annealed for half an hour at about 900$^{\circ}\mathrm{C}$ in a
vacuum better than 6.0$\times 10^{-10}$ Torr. ARPES measurements
were performed at 25 K with an energy resolution of about 40 meV
which mostly stemmed from 0.0075 \AA$^{-1}$ momentum resolution in
combination with the fast dispersion of the bands. The pressure was
better than 5.0$\times 10^{-11}$ Torr. For the experiment, we
exploited the small beam spot ($\approx 40 \mu$m) to locate a flat
region without flakes. A photon energy of 85 eV was used so that
$k_z$ was near \textit{K}. Electronic structure calculation was done
by using the SIESTA code\cite{siesta} based on pseudo-potential
method.

\begin{figure}
\centering \epsfxsize=8.3cm \epsfbox{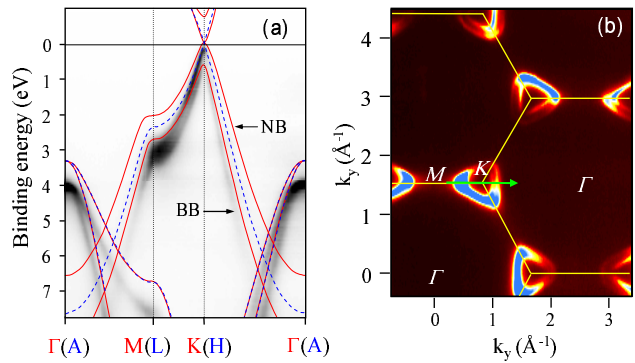} \caption{(a)
Comparison of experimental data and calculated band structure of
graphite along the high symmetry lines.  (b) Intensity map at a
constant binding energy of 1.75 eV. Brillouin zones are drawn as
solid lines.} \label{fig1}
\end{figure}

Fig.\ 1(a) shows the calculated electronic band dispersion of
graphite along the high symmetry lines, $\Gamma$-M-K-$\Gamma$
(solid lines) and A-L-H-A (dashed). Also shown in the figure as an
intensity map is the experimental data with the photon energy
tuned for $k_z=$K. Other than the fact that the calculated band
structure has to be expanded, which is well known\cite{Zhou},
theoretical and experimental results match well. The two bands
near the Fermi energy are $p_z$-derived $\pi$ bands and are split
due to the inter-layer interaction, resulting in anti-bonding
(AB), non-bonding (NB) and bonding (BB) bands. Among them, NB and
BB are occupied and can be seen by ARPES. The splitting $\omega_B$
between the NB and BB is about 0.7$\pm$0.1 eV\cite{Feuerbacher}.
The data can also be plotted in momentum space at a constant
binding energy (1.75 eV) as shown in Fig.\ 1(b). The plot shows an
apparent hexagonal Boulliouin zone expected for graphite. It is
also clearly seen that rounded triangles around the K points are
split. The bands have approximate linear dispersion with
point-like Fermi surfaces at the K points, resulting in an
approximate linear electronic density of states near the Fermi
energy (linear DOS).

Fig.\ 2(a) shows high resolution data taken along the M-K-$\Gamma$
line as marked in Fig.\ 1(b). The two peaks are sharp and clearly
split. We note that there are no defect-induced states as reported
in Ref.\ \cite{Sugawara2}. In addition, the background at high
binding energies is very small, in strong contrast to the
previously reported data \cite{Sugawara, Law, Takahashi, Law2,
Balasubramanian}. We attribute these to the high quality of the
crystals used in our experiment. Fig.\ 2(b) shows the energy
distribution curve (EDC) at the K-point (arrow in panel (a)).  One
peculiar aspect of the data is that the region between the two
peaks are somewhat filled up. As a result, the BB has a tail on
the $lower$ binding energy side contrary to the usual case. This
aspect of the data can be understood as follows. Even though we
tuned the photon energy to the K point, the finite escape depth of
the photoelectron introduces an uncertainty in $k_z$, $\Delta
k_z=1/\mu$ where $\mu$ is the escape depth. This uncertainty in
$k_z$ depicted in Fig.\ 2(c) necessarily brings in broadening due
to the $k_z$ dispersion, which is known as the photoelectron life
time. Note that the NB (BB) has its dispersion maximum (minimum)
at $k_z=0$ and as a consequence the broadening effect will be
one-sided towards the band position at $k_z=\pi/c$ where the two
bands collapse. Fig.\ 2(d) depicts a model spectral function when
all these effects are accounted for. Only when such effects are
considered can one reliably extract the lifetime broadening. We
used $\mu=7${\AA} for the fitting\cite{Tanuma} and the model
function in Fig.\ 2(d) is convolved with a Voigt function with the
Gaussian width set to the total energy resolution of 40 meV. A
typical fit is shown in Fig.\ 2(b) as the solid line.

\begin{figure}
\centering \epsfxsize=8.5cm \epsfbox{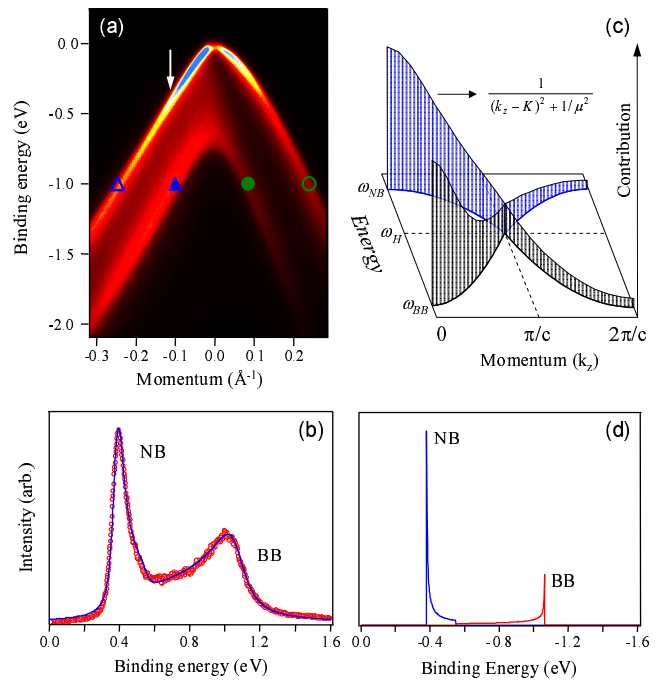} \caption{(a) High
resolution ARPES data taken along the M-K-$\Gamma$ as marked in
Fig.\ 1(b).  (b) A typical EDC (circles), taken at the $k$ value
marked in panel (a) and the fit (line) using the model discussed
in the text.  (c) Schematic of the dispersions along the $k_z$
direction for non-bonding (upper curve) and bonding (lower curve)
bands. Due to the finite escape depth ($\mu$), a range of $k_z$
contributes to the spectral functions.  The contribution
probability of each $k_z$ is represented by the height of the
graph.  (d) Resulting spectral function near K point without
lifetime broadening.} \label{fig2}
\end{figure}

Before we look into the low energy dynamics of graphite, it is
necessary to discuss how the couplings to phonons and other low
energy excitations are affected by the linear DOS of graphite. The
EPC in metallic systems is well understood\cite{Grimvall} and has
been extensively studied recently by using
ARPES\cite{Hengsberger,Valla}. However, there is a relative lack
of understanding of EPCs in semi-metals. In fact, to our best
knowledge, EPC effects on the ARPES lineshape for semi-metals with
linear DOS have not been discussed. As a consequence, models
derived for metals have been improperly used in graphite
studies\cite{Sugawara}.

To understand the EPC in a system with a linear DOS, we evaluate
the Feynman diagram for a single phonon process shown in Fig.\
3(a). We first look at a single band system for simplicity and
will extend the analysis to the graphite case. The contribution
from the process to the imaginary part of the self energy
$\Sigma^{\prime\prime}_{\text{ep}}$ due to EPC at $T$=0
is\cite{Lautenschlager}
\begin{equation}
\Sigma^{\prime\prime}_{\text{ep}}(\omega_k)=\sum_{\nu}\int\bigl|<k^\prime,
q\left|H_1\right|k,0>\bigr|^2f(-\omega_{k^\prime})\delta(\omega_k-
\omega_{k^\prime}-\omega_{\nu,q})dk^\prime
\end{equation} where $k$ and $k^{\prime}$ are
crystal momenta of the holes and $f$ Fermi-Dirac function at
$T=0$, respectively. The phonon momentum is $q=k-k^{\prime}$, and
$\nu$ represents the phonon mode. $H_1$ is the Hamiltonain for the
EPC, and it can be written as
\begin{equation}
H_1=g c^\dagger_{k-q} c_k b^\dagger_q
\end{equation}
where $g$ is the transition probability amplitude for particular
$k$ and $q$. Assuming Einstein phonons and constant $g$,
\begin{equation}
\begin{split}
\int\bigl|<k^\prime , q\bigl|g c^\dagger_{k-q} c_k b^\dagger_q
\bigr|k,0>\bigr|^2\delta(\omega_k-\omega_{k^\prime}-\omega_{\nu,q})dk^\prime
\\=g^2\int \delta(\omega_k-\omega_{k^\prime}-\omega_{\nu,q})dk^\prime=g^2\mathcal{D}(\omega_k-\omega_\nu)
\end{split}
\end{equation}
where $\mathcal{D}$ is the electronic DOS. The expression for
$\Sigma^{\prime\prime}_{\text{ep}}$ then becomes
\begin{equation}
\Sigma^{\prime\prime}_{\text{ep}}(\omega)=\sum_{\nu} g^2
\mathcal{D}(\omega-\omega_\nu)f(\omega_\nu-\omega)
\end{equation}
For a metal, $\mathcal{D}$ is approximately constant in the
immediate vicinity of $E_F$ and above equation reduces to the
expected step function. On the other hand, for a linear DOS,
$\Sigma^{\prime\prime}_{\text{ep}}(\omega)$ increases linearly
beyond the phonon energy as shown in Fig.\
3(b)\cite{Supplementary}.

We may now extend the analysis to the two band case. The EPC
Hamiltonian becomes
\begin{equation}
H_1=\sum_{j}g_{ij} c^\dagger_{k-q,j} c_{k,i} b^\dagger_q
\end{equation}
where $i$ and $j$ are band indices with $i$ representing the
initial band.  $g_{ij}$ with $i=j$ is the EPC for intra-band
transition while $g_{ij}$ with $i\neq j$ for inter-band
transition.  The result is shown schematically in Fig.\
3(c)\cite{Supplementary}. Note that
$\Sigma^{\prime\prime}_{\text{ep}}$ for the NB has an upward kink
at the energy where the BB starts. The kink and slopes of the self
energy curves carry the information on the EPC matrix element
$g_{ij}$.

The difficulty in extracting the EPC constant $\lambda$ comes from
the fact that it is almost impossible to $guess$ the bare bands.
In metals, the bare band can be approximated by a linear band
(possibly with an additional small parabolic term) over the energy
window of interest which can be obtained by fitting the
experimental dispersion.  This method, however, cannot be used for
semi-metals and insulators for an obvious reason.  In fact, the NB
band of graphite near K does not form a Dirac cone but has a
fairly strong curvature as is well known from the band
calculation.  Therefore, the standard method of obtaining the real
part of the self energy $\Sigma^\prime$ by taking the difference
between the bare band and the experimental dispersion can not be
used. Alternatively, $\Sigma^\prime$ can be obtained from
$\Sigma^{\prime\prime}$ through the Kramers-Kronig relation.  Then
the EPC constant $\lambda$ can be extracted from the real part of
the self energy using the following formula\cite{Grimvall}.
\begin{equation}
\lambda=-\frac{\partial\Sigma^\prime_{\text{ep}}(\omega)}{\partial\omega}\biggr|_{\omega=0}
\end{equation}
We point out that this method usually requires higher quality of
data compared to the standard method to extract the similar
quality of information.

\begin{figure}
\centering \epsfxsize=7.5cm \epsfbox{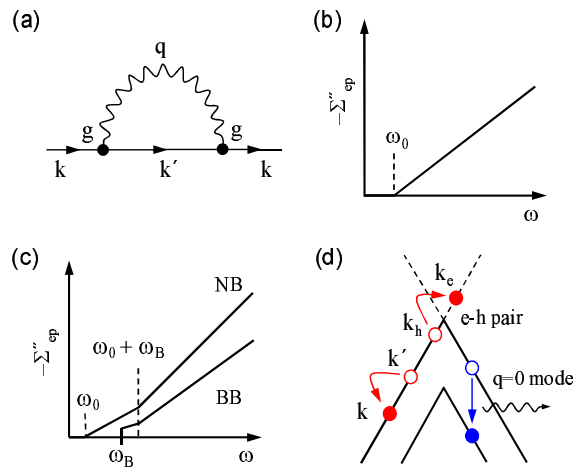} \caption{(a)
Feynman diagram for the EPC under consideration.  (b) Schematic of
the $\Sigma^{\prime\prime}$ for the phonon decay for a single band
system with a linear DOS. (c)$\Sigma^{\prime\prime}$ for a double
band system.  (d) Other possible decay channels for photo-holes,
red for decay through electron-hole creation and blue for $q$=0
mode decay. Note that $k^\prime$ and $k_h$ can be changed for the
electron-hole creation case.} \label{fig3}
\end{figure}

Finally, we discuss other decay channels (see Fig.\  3(d)).  The
photo-hole may decay by creating an electron-hole pair or by
exciting a $q$=0 mode in addition to emitting a phonon.  For the
decay through electron-hole pair creation, the energy and momentum
conservations mandate
$\Delta\omega$=$\omega_{k^\prime}-\omega_k$=$\omega_e-\omega_h$
and
$\Delta\textbf{k}$=$\textbf{k}^\prime$-$\textbf{k}$=$\textbf{k}_{e}$-$\textbf{k}_{h}$.
Note that both $\textbf{k}^\prime$ and $\textbf{k}$ are in the
same (lower) Dirac cone while $\textbf{k}_{e}$ and
$\textbf{k}_{h}$ are in different Dirac cones. In such case, for
the equations to be met, both $\textbf{k}^\prime$-$\textbf{k}$ and
$\textbf{k}_{e}$-$\textbf{k}_{h}$ vectors should lie in the
steepest decent line on the 2 dimensional conical dispersion (the
configuration illustrated in the figure) and the available phase
space volume for the transition in
$\Delta\omega$-$\Delta\textbf{k}$ space is
zero\cite{Supplementary}. For the $q=0$ mode decay, the
contribution is considered to be photo-hole momentum-independent
as the two bands are almost parallel to each other.  Note that NB
band is not susceptible to $q=0$ mode decay as one can see in the
figure, giving a constant upward shift only to
$\Sigma^{\prime\prime}$ of the BB.

\begin{figure}
\centering \epsfxsize=8.5cm \epsfbox{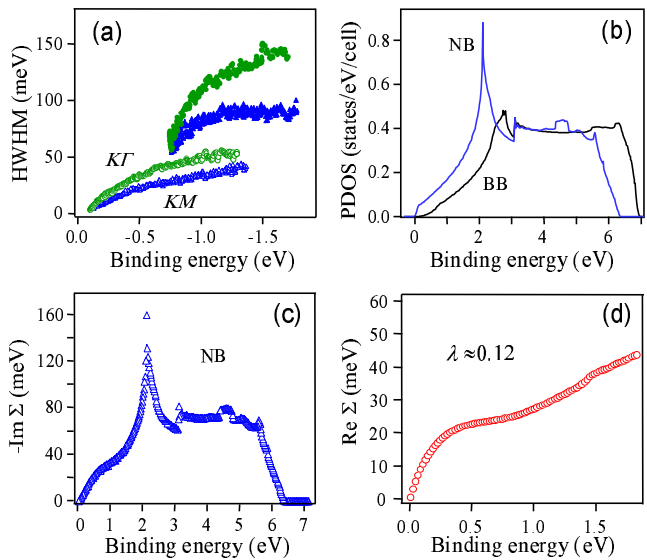} \caption{(a)
$\Sigma^{\prime\prime}$ (half of the Lorentzian width) \emph{vs.}
binding energy.  (b) Calculated PDOS of the NB and BB. (c)
$\Sigma^{\prime\prime}$ for the NB over the entire energy range
range by combining the experimental data and the calculated PDOS.
(d) $\Sigma^\prime$ obtained by Hilbert transforming the
$\Sigma^{\prime\prime}$ in panel (c).} \label{fig4}
\end{figure}

With the finite escape depth considered, the half width of the
Lorentzian is extracted and plotted in Fig.\ 4(a).  Filled and empty
symbols represent the BB and NB, respectively as marked in Fig.\
2(a).  Overall, the width monotonically increases without particular
features.  There are smooth slope changes at around 0.4 eV which
roughly coincide with energy where $k_z$ dispersion of the NB
saturates.  Therefore, we attribute this to the yet unaccounted band
structure (in combination with the finite escape depth)
effect\cite{BandEffect}.  When such an effect is considered through
additional modeling, we expect the width will increase more
linearly.  While the values for the NB, when converted to life time,
are reasonably compatible with the values obtained by time resolved
techniques\cite{Xu,Moos}, they are much smaller than the previously
reported values\cite{Sugawara}.

From the measured widths, we find the the following additional
facts.  Firstly, the fact that the observed widths do not show a
high order behavior ($\sim\omega^2$) reveals that the EPC is the
dominant decay channel for the photo-holes.  This is expected
because the available phase space is very small as discussed
above.  Secondly, no (or very weak at most) kinky feature at the
optical phonon energy ($\sim 0.2$ eV) is observed.  It was
previously reported that a strong EPC was observed in the
imaginary part\cite{Sugawara} of the self energy.  Our observation
shows it is not the case (at least near the K point). Thirdly, the
inter-band decay, if exists, is much weaker than the intra-band
evidenced by the lack of an upward kink in the NB band curve at
the onset energy of BB (see Fig.\ 3(c)).  This allows us to have a
convenience of treating the NB and BB independently.  As will be
clear, this makes the task of calculating the EPC constant much
simpler.  Fourthly, as the inter-band decay is very small, the
width of the BB at the onset energy can not be explained by the
EPC. It has to come from another mechanism such as the $q=0$ mode
decay as discussed earlier which is not well-understood. Lastly,
the slopes of the curves are somewhat $k$-dependent (that is, the
slopes for the K-M cut are steeper).  This $k$-dependence suggests
that there is $k$-dependent EPC in the system.  Such behavior is
indeed predicted theoretically\cite{Spataru} and is also observed
in the study of a single layer graphene\cite{Jessica}.

Even though the constant offset in the $\Sigma^{\prime\prime}$ of
the BB is not well-understood, one can estimate the EPC of the NB
as the inter-band transition is weak.  As discussed above,
calculating the EPC constant $\lambda$ requires knowledge of
$\Sigma^{\prime\prime}$ over the entire energy range.  As it is
not available, we approximated $\Sigma^{\prime\prime}$ at high
energies by the scaled, calculated partial DOS (PDOS) to match the
the experimental data, noting that $\Sigma^{\prime\prime}$ is
proportional to PDOS as in equation (3).  The resulting
$\Sigma^{\prime\prime}$ for the NB along the $K$-$M$ direction is
plotted in panel (c).  $\Sigma^\prime$ is then obtained by a
Hilbert transformation of the symmetrized $\Sigma^{\prime\prime}$
and plotted in panel (d). The $\lambda_{NB}$ value obtained from
$\Sigma^\prime$ is about 0.12, which is much smaller than the
previously estimated value from ARPES\cite{Sugawara}.  For the
K-$\Gamma$ direction, $\lambda$ was 0.2. Therefore, we set 0.2 as
the upper bound for EPC constant $\lambda$.  This value is very
small compared to $\lambda$ values in typical metals as expected
for semi-metals with very small DOS at the Fermi energy.

Authors acknowledge helpful discussions with J. H. Han. This work
is supported by the KICOS through a grant provided by MOST in
M60602000008-06E0200-00800, by KOSEF through CSCMR, and also by
MOST through GPP. ALS is operated by the DOE's Office of BES.


\begin{thebibliography}{25}
\bibitem{BandEffect} To be precise, the Lorentzian width of the Voigt function
should be varied as a function of energy in convolving the model
function. That is, one has to do energy dependent convolution for
each EDC.
\end{thebibliography}
\end{document}


This file contains supplementary information on the the self
energies due to the electron-phonon coupling (EPC) and electron
hole pair creation in graphene and graphite which have
characteristic linear electronic density of states near the Fermi
energy (linear DOS). We first investigate the scattering process
through phonons by using schematic diagrams for the single
(graphene) and double (graphite) conical band structures. It
allows one to intuitively understand how characteristic linear DOS
contributes to the scattering rate of photo-hole. Then we discuss
the scattering process by electron-hole pair creation by
considering the available phase space for the process. We show
that this process gives little contribution to the scattering rate
of a photo-hole.

\section{Scattering rate due to electron-phonon coupling}

\subsection{Single band case}
In this case, we assume a conical band structure with the Fermi
energy at the apex of the cone. There also is an unoccupied
conical band above the Fermi energy, forming a Dirac-cone-like
band structure as shown in Fig. 1(a). Assume an Einstein phonon at
energy $\omega_0$. If a photo-hole with momentum $\mathbf{k}$
(filled circle) is created by a photon as shown in Fig. 1(a), it
can be filled by an electron with energy of
$\omega_{k^\prime}=\omega_k -\omega_0$ and momentum
$\mathbf{k}^\prime=\mathbf{k}-\mathbf{q}$ (empty circles) where
$\mathbf{q}$ is the phonon momentum. We further assumed for
simplicity the scattering amplitude $g$ is momentum independent.
In that case, the scattering rate is proportional to the number of
such $\mathbf{k}^\prime$ states, thus the DOS at $\omega_k
-\omega_0$. Note that if the binding energy of the photo-hole is
smaller than the phonon energy $\omega_0$, the scattering cannot
occur because there are no electrons with sufficient energy to
emit a phonon. Therefore, the self energy of photo-hole as a
function of the binding energy is proportional to
$\mathcal{D}(\omega_k-\omega_0)$ and looks like the schematic
shown in Fig. 1(b).

\begin{figure}
\centering \epsfxsize=8.3cm \epsfbox{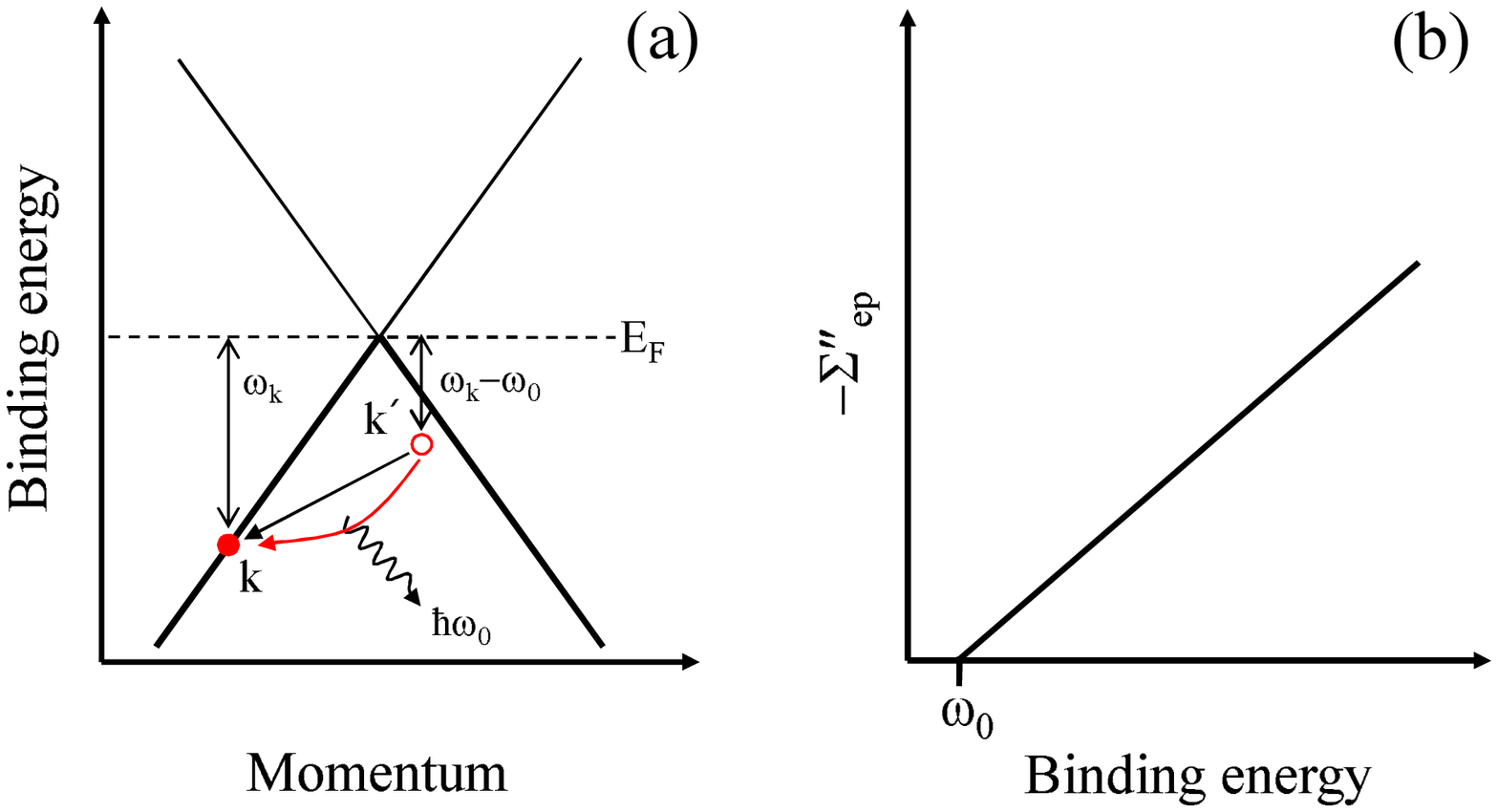} \caption{(a)
Schematic diagram of the photo-hole decay by emitting a phonon in
the single band case. $\omega_0$ is phonon energy and $\mathbf{k}$
and $\mathbf{k}^\prime$ are initial and final photo-hole states,
respectively. (b) Schematic of the resulting decay rate (the
imaginary part of self energy) vs binding energy. It is zero up to
$\omega_0$ and then increases linearly.} \label{fig1}
\end{figure}

\begin{figure}
\centering \epsfxsize=8.3cm \epsfbox{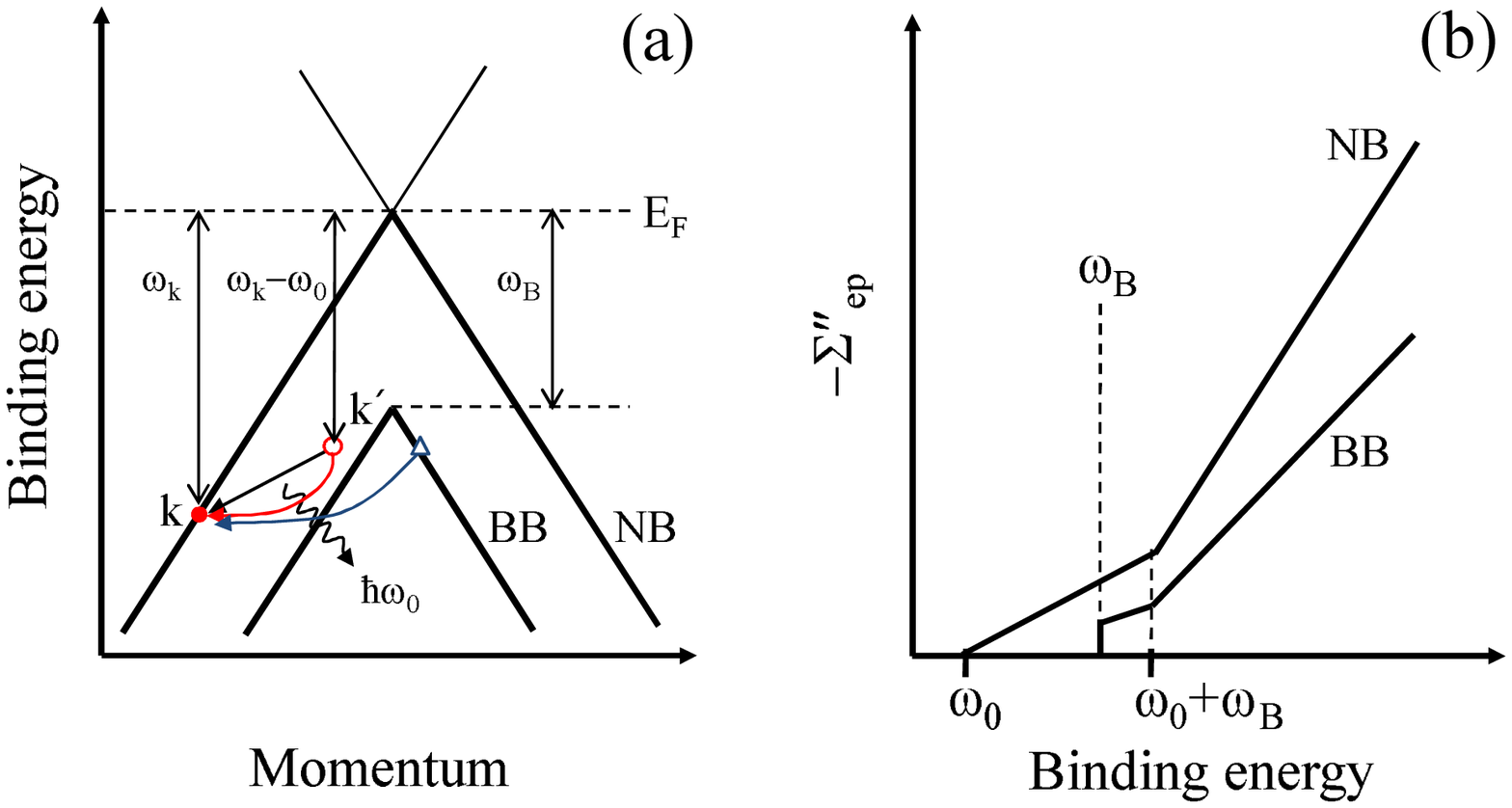} \caption{(a)
Schematic diagram of the photo-hole decay in the double band case.
$\omega_B$ is the energy of the apex of the lower conical band.
(b) The imaginary parts of the self energies vs binding energy for
photo-holes created in non-bonding and bonding bands.}
\label{fig2}
\end{figure}

\subsection{Double band case}
For the double band case, another conical band exists $\omega_B$
below the Fermi energy as shown in Fig. 2(a). One needs to
investigate the cases where the initial photo-hole is in
non-bonding and bonding bands separately. We again count the the
number of available electrons that can fill the photo-hole. We
first consider the case in which the photo-hole is created in
non-bonding band. If the binding energy of the photo-hole is
smaller than $\omega_B +\omega_0$, the decay rate (the number of
electrons that can fill the photo-hole) will be the same as that
in the single band case. When the binding energy becomes larger
than $\omega_B +\omega_0$, the additional electrons (triangles in
Fig. 2(a)) in the second band begin to contribute to the
scattering process. This additional contribution is proportional
to the partial DOS of the second band at $\omega_k-\omega_B$ and
thus linearly increases. Note that the additional slope of the
linear increase depends on the inter-band scattering amplitude
$g_{ij}$ ($i\neq j$). The resulting decay rate is schematically
shown in Fig. 2(b). We now investigate the case for a photo-hole
in the bonding band. If the energy of the photo-hole is larger
than $\omega_B$ but smaller than $\omega_B +\omega_0$, the
photo-hole can be filled with only electrons in non-bonding band
via inter-band transition. Meanwhile the photo-hole can decay via
inter- and intra-band transitions when the energy is larger than
$\omega_B +\omega_0$. The resulting decay rate is also
schematically shown in Fig. 2(b). If the electron-phonon couplings
are different for bonding and non-bonding bands, the overall
slopes of self energies will be different as shown in Fig 2.(b).

\section{Available phase space for electron-hole pair creation}
A photo-hole with the momentum $\mathbf{k}$ is created in the
Dirac cone as shown in Fig 3.(a). It may go through a transition
to a $\mathbf{k^\prime}$ state by creating an electron-hole pair
(denoted as $\mathbf{k_e}$ and $\mathbf{k_h}$, respectively in Fig
3.(a)). If we plot the energy difference
($\Delta\omega$=$\omega_k-\omega_{k^\prime}$) as a function of
momentum difference
$\Delta\mathbf{k}$=$\mathbf{k}-\mathbf{k}^\prime$, the possible
transitions occupy the lower right of the
$\Delta\omega$-$\Delta\mathbf{k}$ phase space as shown in Fig.
3(b) (grey). In a similar way, we can go through the same process
for the electron-hole pair creation. One can show that
$\Delta\omega$=$\omega_e-\omega_h$ and
$\Delta\mathbf{k}$=$\mathbf{k_e}-\mathbf{k_h}$ of the
electron-hole creation processes occupy the upper left of the
phase space (hatched). We note that only the boundary is shared by
two regions. Energy and momentum conservations enforce the decay
through the electron-hole pair creation to occur in the phase
space points that belong to both regions. Therefore, the available
phase space volume for the decay is zero and thus the transition
rate.

\begin{figure}
\centering \epsfxsize=8.3cm \epsfbox{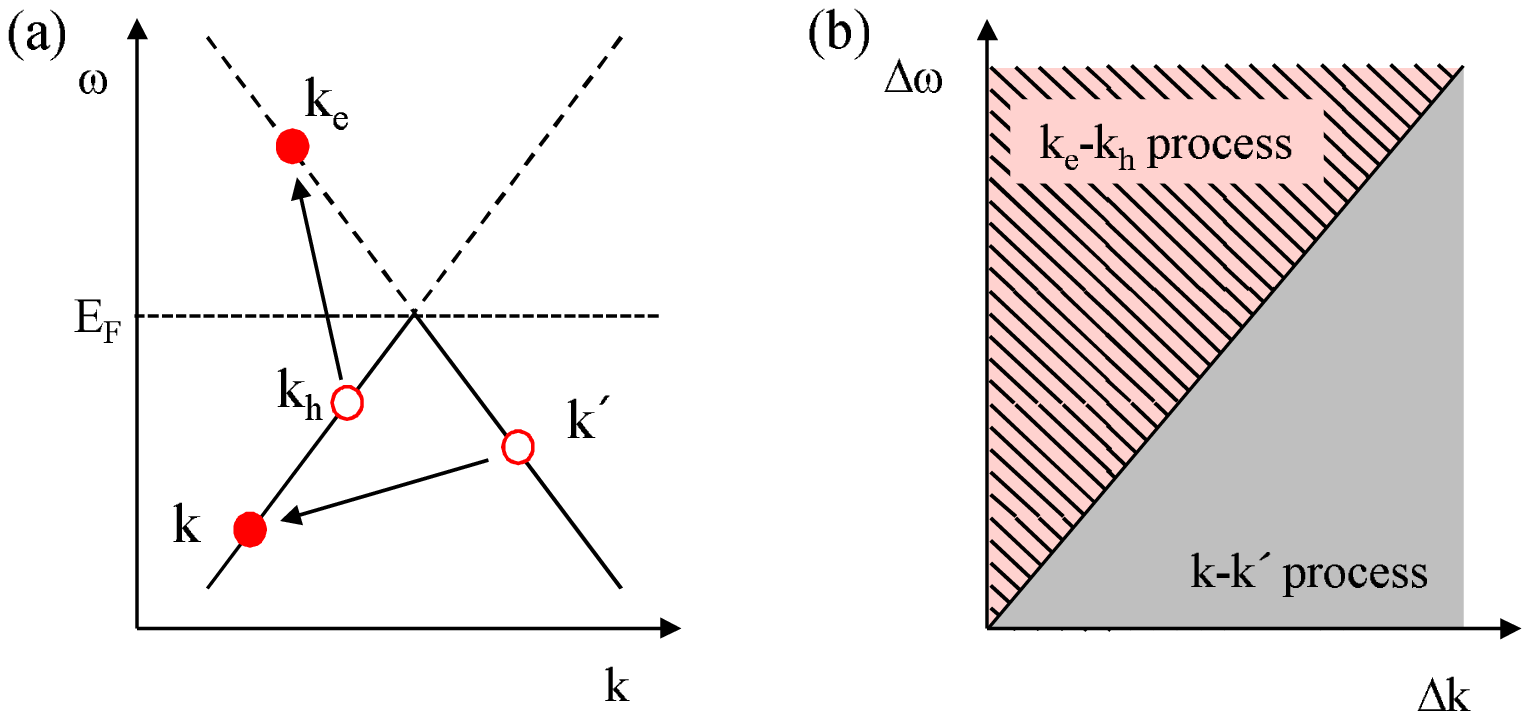} \caption{(a)
Schematic of electron-hole creation($\mathbf{k}_e -\mathbf{k}_h$)
and transition of a photo-hole ($\mathbf{k}_e -\mathbf{k}^\prime$)
in a conical band structure. (b) Available phase spaces for
$\mathbf{k}_e -\mathbf{k}_h$ (hatched region) as well as
$\mathbf{k}_e -\mathbf{k}^\prime$ (grey region) processes. Two
regions overlap only at the boundary.} \label{fig3}
\end{figure}